\documentclass[journal]{IEEEtran}
\usepackage[T1]{fontenc}
\usepackage[latin9]{inputenc}
\usepackage{amssymb}
\usepackage{graphicx}
\usepackage[unicode=true,
 bookmarks=true,bookmarksnumbered=true,bookmarksopen=true,bookmarksopenlevel=1,
 breaklinks=false,pdfborder={0 0 0},pdfborderstyle={},backref=false,colorlinks=false]
 {hyperref}
\hypersetup{pdftitle={Your Title},
 pdfauthor={Your Name},
 pdfpagelayout=OneColumn, pdfnewwindow=true, pdfstartview=XYZ, plainpages=false}

\makeatletter
\usepackage[caption=false,font=footnotesize]{subfig}

\makeatother

\begin{document}
\title{Antenna Coding by Pixel Antennas: Unlocking A New Dimension for Wireless
Communications}
\author{Zixiang Han,~\IEEEmembership{Member,~IEEE,} Shanpu Shen,~\IEEEmembership{Senior Member,~IEEE,}
Yifei Yuan,~\IEEEmembership{Fellow,~IEEE,}\\
Jiangzhou Wang,~\IEEEmembership{Fellow,~IEEE,} Ross Murch,~\IEEEmembership{Fellow,~IEEE,}~and~Chih-Lin
I,~\IEEEmembership{Life Fellow,~IEEE}\thanks{This work was supported by China Mobile Research Institute, and the
Science and Technology Development Fund, Macau SAR (File/Project no.
001/2024/SKL, 0002/2025/EQP and 0078/2026/RIB1). \textit{(Corresponding
author: Shanpu Shen)}}}
\maketitle
\begin{abstract}
Conventional wireless communication systems employ antennas with fixed
radiation patterns. Enhancing system performance by increasing the
number of antennas would push the limits of cost and size for communication
devices. To overcome this issue, novel antenna technologies are urgently
desired to break through the fundamental performance bottleneck and
engineering limits. Pixel antennas provide a general framework for
highly reconfigurable antennas, which can flexibly adjust radiation
patterns via controlling variable lumped elements placed across adjacent
pixels, unlocking novel degrees of freedom (DoF) in the beamspace
domain. Recently, an emerging antenna coding technique links wireless
system performance to the antenna configurations. This enables the
use of pixel antennas to enhance system performance by leveraging
the DoF from the antenna side. In this article, we provide a comprehensive
overview on antenna coding by pixel antennas. Specifically, based
on the pixel antenna model, the concept of antenna coding is firstly
introduced, followed by a discussion on the classifications of antenna
coding. Two cases for channel gain and multiplexing gain enhancement
are highlighted to show the effectiveness of antenna coding in wireless
communications. Besides, various optimization approaches for antenna
coding design are investigated. Finally, the challenges and opportunities
are outlined, providing guidance for future works.
\end{abstract}

\section{Introduction}

The evolution of wireless communications has significantly changed
the way of data transmission over the past few decades. A key enabler
of capacity improvement is the multiple-input multiple-output (MIMO)
technique. From third-generation small-scale MIMO to fifth-generation
massive MIMO, the number of antennas has grown dramatically to support
the increasing demand for higher data rates. However, this growth
has simultaneously driven up cost, size, weight, and power consumption
of wireless equipment, all of which have become the bottleneck for
further improving data rates in future sixth-generation (6G) networks.
Simply increasing antennas is no longer sufficient to meet the performance
requirement of 6G communications. Therefore, new technologies are
urgently needed to provide additional degrees of freedom (DoF) beyond
spatial, time, and frequency domains.

\begin{figure*}[t]
\begin{centering}
\textsf{\includegraphics[width=12cm]{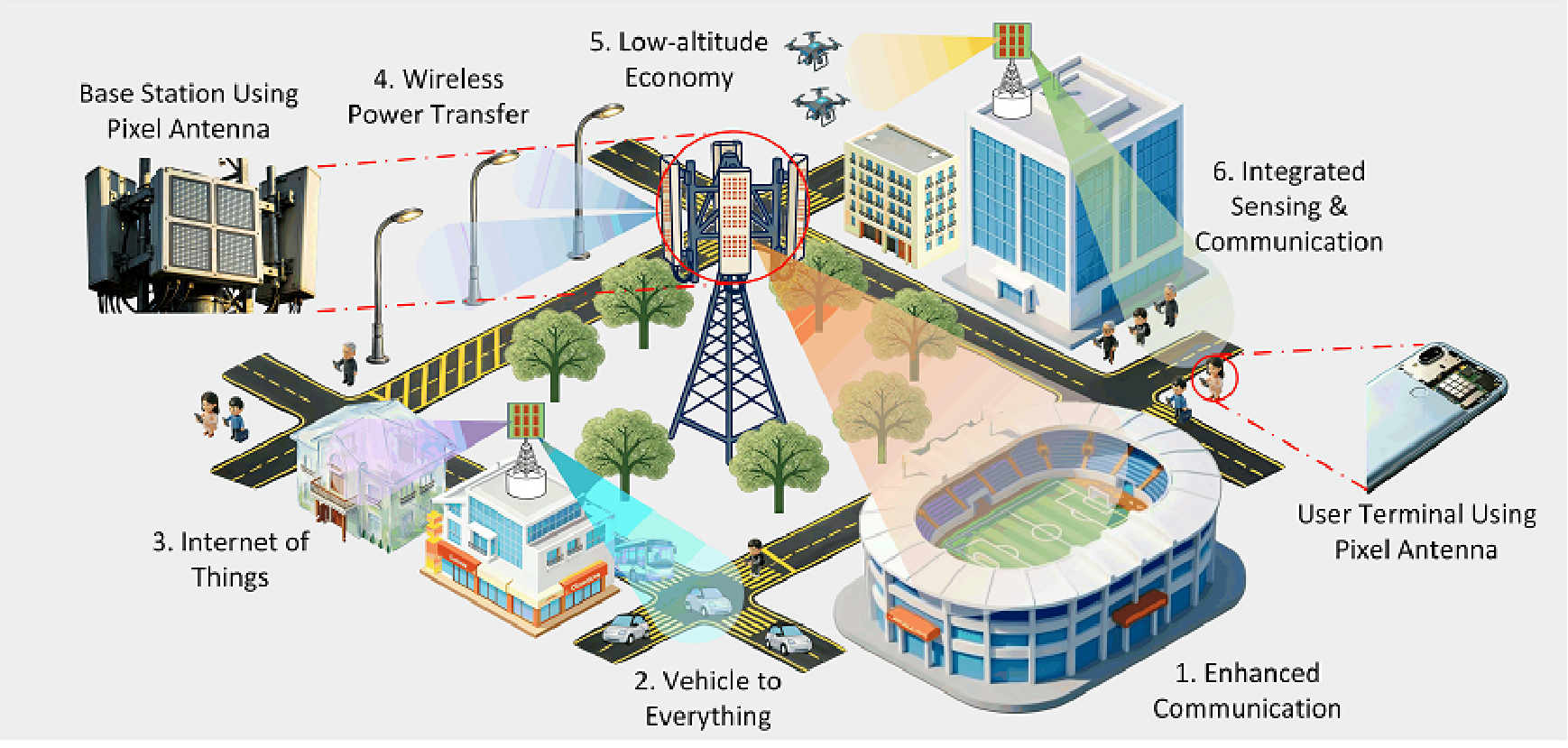}}
\par\end{centering}
\caption{Applications of pixel antennas in various scenarios. \label{fig:Scenario}}
\end{figure*}
Reconfigurable antennas provide a promising pathway to introduce DoF
from the antenna side. By dynamically tuning their configurations,
these antennas can adapt to varying channel conditions or system requirements.
This reconfigurability transforms the antenna configuration into a
new DoF. Pixel antennas represent a general framework for implementing
highly reconfigurable antennas since arbitrary antenna radiating structures
can be discretized into a grid of pixels \cite{1367557}. Adjacent
pixels can be interconnected via variable lumped elements. The pixel
antenna topology is reconfigured by controlling the direct current
(DC) biased voltages imposed on these elements, enabling diverse applications,
including reconfigurable intelligent surfaces (RIS), wireless power
transfer, and array decoupling. Therefore, pixel antennas can be deployed
in various scenarios as shown in Fig. \ref{fig:Scenario}. The recently
emerged fluid antenna system (FAS) is a technique that links antenna
reconfigurability to wireless system performance \cite{11302793},
\cite{11513728}, while it does not specify the type of reconfigurable
antenna so that various approaches are possible. Pixel antennas have
been preliminarily used in FAS where the topology reconfiguration
mimics antenna movement and geometry changes in FAS \cite{11263876}.
However, unlocking the full potential of pixel antennas to enhance
wireless communications remains a key challenge.

\begin{figure*}[t]
\begin{centering}
\textsf{\includegraphics[width=2.2cm]{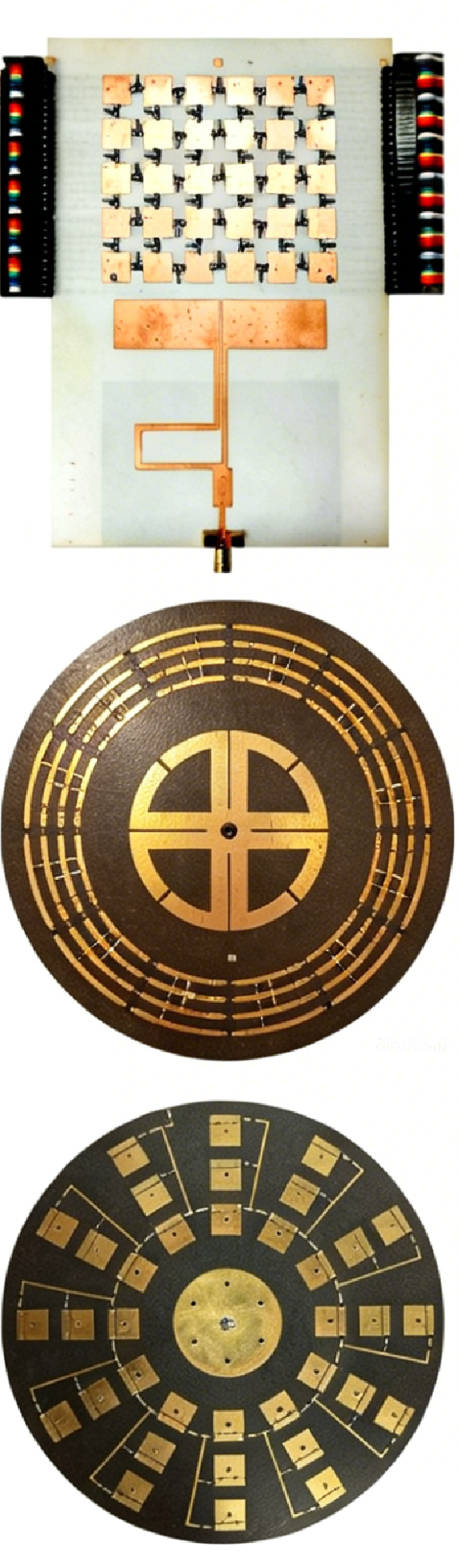}$\quad\;$\includegraphics[width=2.2cm]{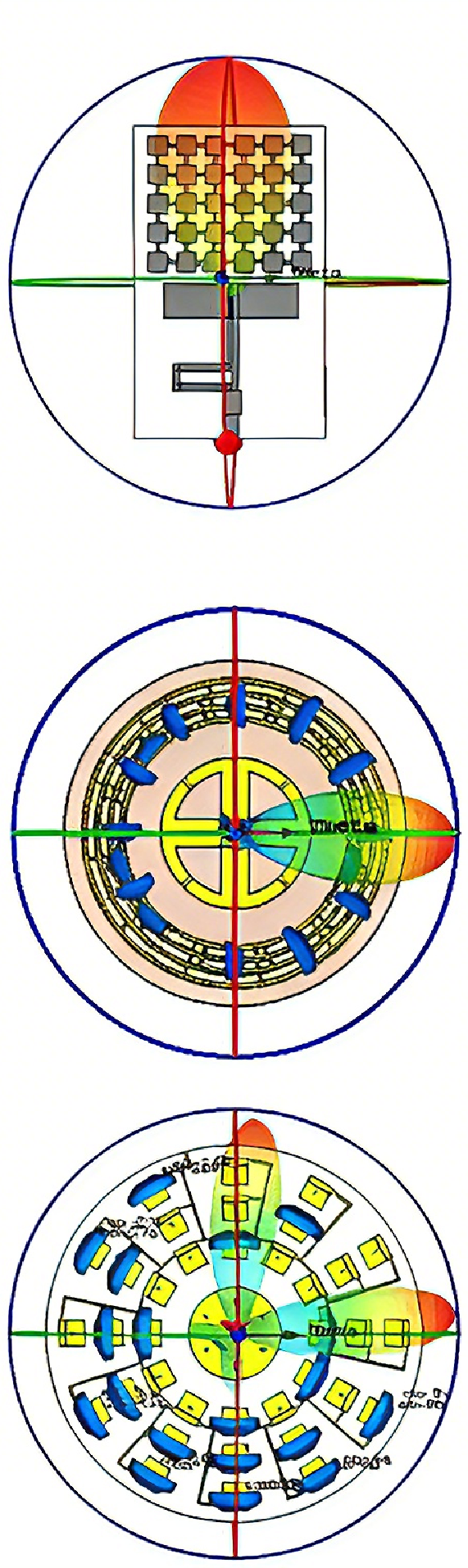}$\quad\;$\includegraphics[width=2.2cm]{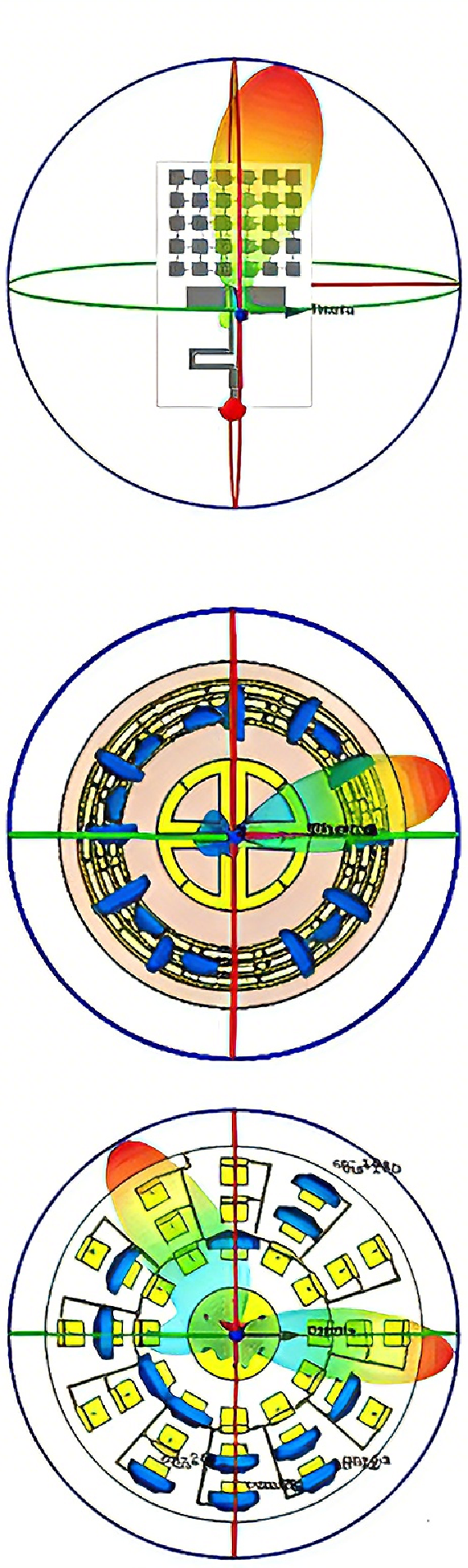}$\quad\;$\includegraphics[width=2.2cm]{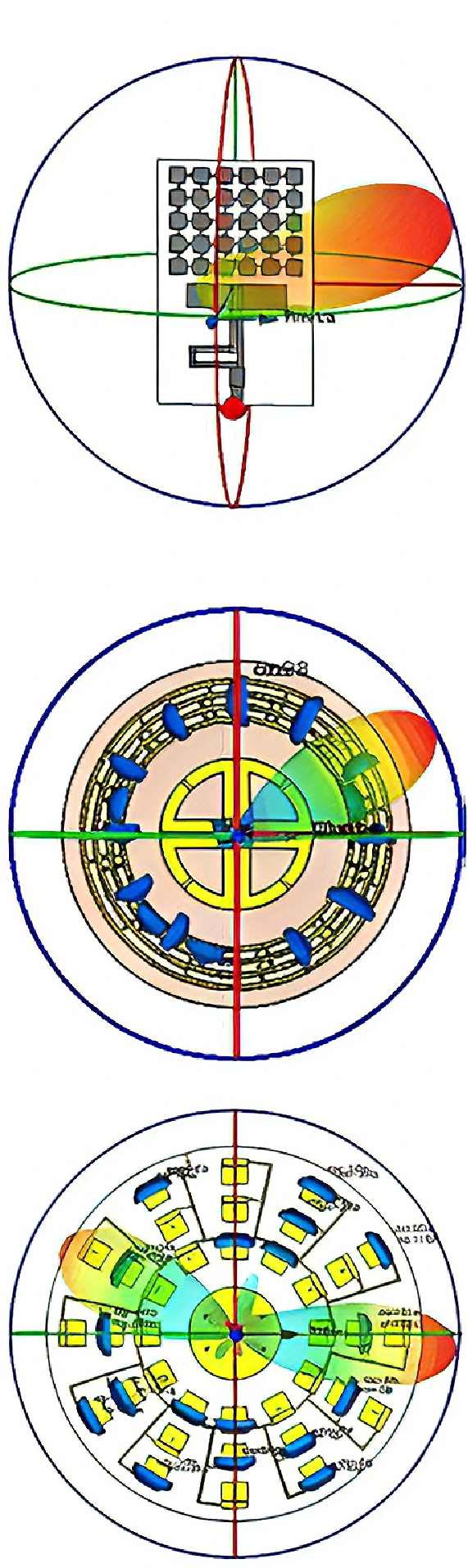}$\quad\;$\includegraphics[width=2.2cm]{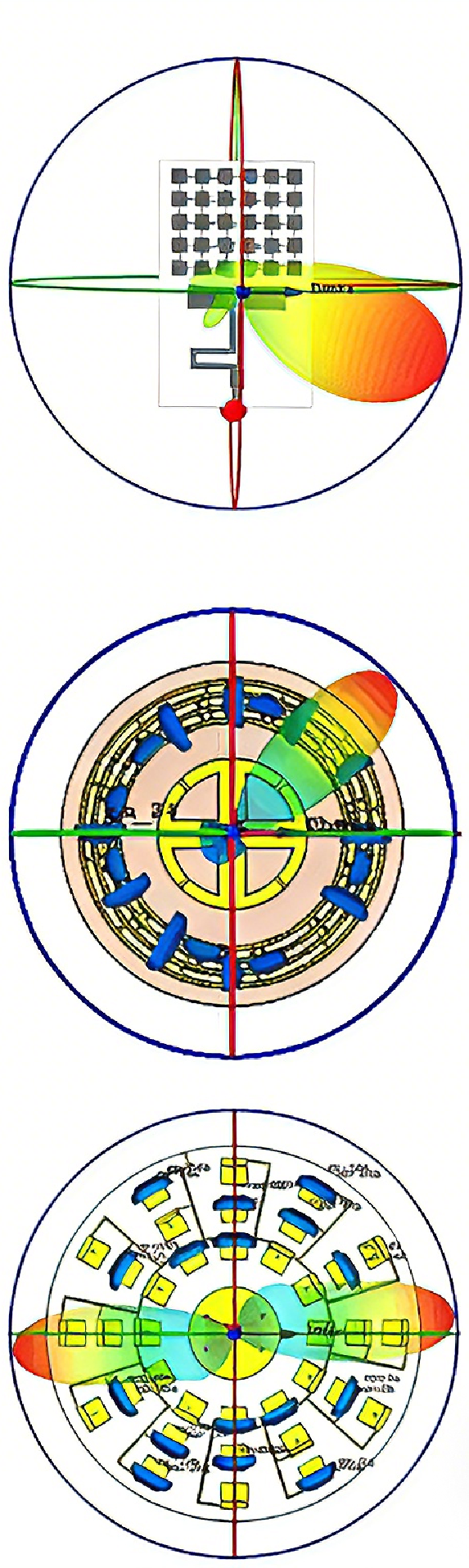}}
\par\end{centering}
\caption{Various pixel antenna designs {[}7{]}-{[}9{]} (first column) and corresponding
radiation patterns (second to fifth columns). \label{fig:Pixel Antenna}}
\end{figure*}
To overcome this challenge, a technique called antenna coding has
been developed \cite{shen2024antenna}. In this technique, electromagnetic
(EM) characteristics of pixel antennas become programmable by using
antenna coding to control the states of variable lumped elements.
Based on the beamspace channel representation, the system model can
be expressed as a function of the antenna coder to exploit radiation
pattern reconfigurability when designing wireless communications.
Utilizing antenna coding, channel gain, spectral efficiency (SE) and
energy efficiency can be greatly improved by leveraging the DoF of
antenna reconfiguration \cite{Han2026}.

This article provides a comprehensive overview on the antenna coding
technique. Antenna coding can be classified according to variable
lumped elements and switching interval. Then two key cases are studied.
The optimization approaches and technical challenges are also discussed.

\section{Pixel Antenna}

The concept of the pixel antenna is based on discretizing a continuous
radiation surface of arbitrary shape into a set of individual sub-wavelength
elements, referred to as pixels, and connecting adjacent pixels through
variable lumped elements such as RF switches. Pixel antennas provide
a general framework for reconfigurable antennas since antenna radiating
structures of various topologies can be implemented by tuning variable
lumped elements to adjust the connections between adjacent pixels.
Thus, pixel antennas provide massive reconfigurability to flexibly
control antenna characteristics including operating frequency, radiation
patterns, and polarizations.

As illustrated in Fig. \ref{fig:Pixel Antenna}, various examples
of pixel antennas can be implemented by discretizing Yagi-Uda antennas,
dipole antennas and patch antennas into pixels \cite{lotfi2017printed}-\nocite{Zhang2022a}\cite{Zhang2022},
where adjacent pixels are connected via PIN diodes. The configuration
of pixel antennas is controlled by switching ON or OFF the PIN diodes
via DC biased voltages, so that surface current distribution on pixel
antennas can be altered and the resulting radiation pattern can be
flexibly shaped. Thus, by optimizing the states of PIN diodes, the
desired radiation patterns can be generated, for example steering
the beam toward different directions.

Given the high reconfigurablity enabled by pixel antennas, it is important
to investigate how to control variable lumped elements so that such
reconfigurablity can be leveraged to enhance wireless communications
at the system level.

\section{Antenna Coding}

In this section, we present the emerging technique denoted as antenna
coding, which controls variable lumped elements to adjust antenna
characteristics, such as radiation patterns, to enhance wireless systems.

\subsection{Multiport Network Model}

\begin{figure}[t]
\begin{centering}
\textsf{\includegraphics[width=8cm]{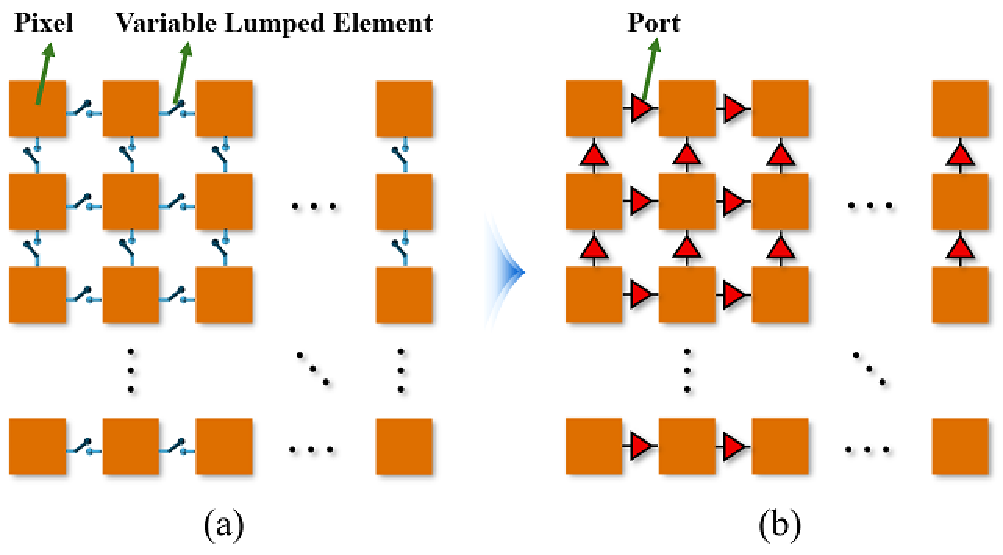}}
\par\end{centering}
\begin{centering}
\textsf{\includegraphics[width=6.5cm]{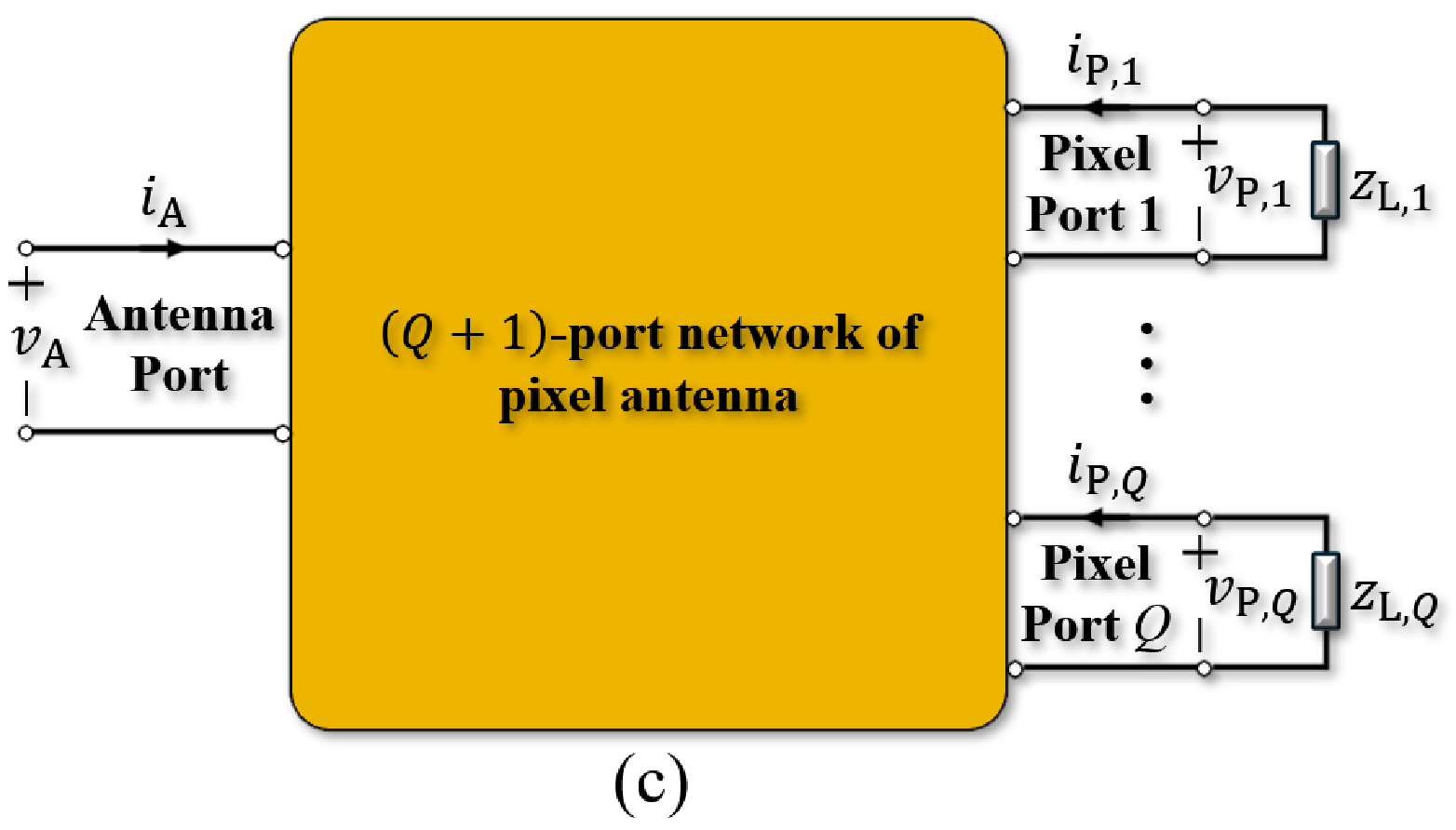}}
\par\end{centering}
\caption{Illustrative schematic for (a) the pixel antenna, (b) the corresponding
multiport circuit network and (c) the equivalent circuit model. \label{fig:Model}}
\end{figure}

A pixel antenna, as shown in Fig. \ref{fig:Model}(a), embedded with
$Q$ variable lumped elements can be modeled as a $\left(Q+1\right)$-port
network, as shown in Fig. \ref{fig:Model}(b), by utilizing $Q$ pixel
ports to replace $Q$ variable lumped elements. The $\left(Q+1\right)$-port
network contains a single antenna port connected to an RF chain and
$Q$ pixel ports loaded with $Q$ variable lumped elements. A variable
lumped element can be modeled as load impedance $z_{\mathrm{L},q}=jx_{\mathrm{L},q}$,
$\forall q\in\mathcal{Q}=\left\{ 1,...,Q\right\} $, where $x_{\mathrm{L},q}$
is the variable reactance. To represent the states of $Q$ variable
lumped elements, a vector $\mathbf{b}=\left[b_{1},...,b_{Q}\right]$
with $b_{q}\in\left[-1,1\right]$, $\forall q\in\mathcal{Q}$ is defined,
which is referred to as \textit{antenna coder}. The antenna coder
is uniquely mapped to the reactance of $Q$ variable lumped elements
through $x_{\mathrm{L},q}=x_{\mathrm{oc}}b_{q}$, $\forall q\in\mathcal{Q}$,
where $x_{\mathrm{oc}}>0$ is very large such as $x_{\mathrm{oc}}=10^{9}$
for analytical purpose. Thus, the positive and negative values of
$b_{q}$ are mapped to capacitive and inductive reactances respectively.
Particularly, $b_{q}=0$ is mapped to zero reactances. indicating
a short circuit. While $b_{q}=-1$ and $b_{q}=1$ are mapped to very
large capacitive and inductive reactances, respectively, both of which
can numerically approximate the open circuit. When evaluating practical
pixel antenna system performance, the parasitic effects of variable
lumped elements can be incorporated into the load impedance $z_{\mathrm{L},q}$.

Based on multiport network analysis \cite{6678331}, with the knowledge
of impedance matrix and open-circuit radiation patterns obtained by
one-time EM simulation, we can formulate the closed-form expression
of antenna characteristics, such as current distribution and radiation
patterns, as a function of the antenna coder. Therefore, using the
multiport network model as shown in Fig. \ref{fig:Model}(c), characteristics
of pixel antennas with arbitrary antenna coders can be efficiently
computed in closed-form expressions. This is significantly more efficient
than using EM solver to simulate pixel antennas with different antenna
coders, and allows optimizing antenna coding in real time to enhance
wireless system performance.

\subsection{Beamspace Channel Representation}

To incorporate the reconfigurable radiation patterns of pixel antennas
into typical wireless system models, the beamspace channel representation
is adopted \cite{sayeed2002deconstructing}. In the beamspace channel
model, the channel between transmit and receive antennas is represented
by the superposition of multiple paths weighted by radiation patterns
of transmit and receive antennas. For example, considering a single-input
single-output (SISO) system where radiation patterns of transmit and
receive antennas are denoted as $\mathbf{e}_{\mathrm{T}}$ and $\mathbf{e}_{\mathrm{R}}$
respectively, the SISO channel in beamspace can be written as $h=\mathbf{e}^{T}_{\mathrm{R}}\mathbf{H}_{\mathrm{V}}\mathbf{e}_{\mathrm{T}}$,
where $\mathbf{H}_{\mathrm{V}}$ is the virtual channel matrix collecting
channel gains of multiple paths, from each angle of departure to each
angle of arrival. Hence, by performing antenna coding, we can dynamically
control radiation patterns of pixel antennas to enhance wireless systems.

\subsection{Orthogonal Basis Radiation Patterns}

To systematically analyze the beamspace channel and get insights into
the fundamental limits of pixel antennas, we consider the orthogonal
basis radiation patterns of pixel antennas, which can be found by
different approaches:
\begin{enumerate}
\item \textbf{Singular Value Decomposition (SVD)} \cite{shen2024antenna}:
Using SVD to decompose the matrix collecting open-circuit radiation
patterns of the $(Q+1)$-port network for the pixel antenna. The unitary
left singular matrix directly provides the orthogonal basis radiation
patterns.
\item \textbf{Theory of Characteristic Modes (TCM)} \cite{mautz1973modal}:
TCM decomposes the radiation pattern of an $N$-port structure into
multiple orthogonal characteristic modes by solving a weighted eigenvalue
problem. The solved eigenvectors are the modal currents that can excite
the orthogonal basis radiation patterns. However, TCM is only applicable
to lossless $N$-port structure and 3-dimenional uniform power angular
spectrum (PAS).
\item \textbf{Pattern Correlation Decomposition Method (PCDM)} \cite{PCDM}:
PCDM can find the orthogonal basis radiation patterns of any $N$-port
structure (lossy or lossless) with arbitrary PAS by performing eigenvalue
decomposition (EVD) on the correlation matrix of open-circuit radiation
patterns. The orthogonal basis radiation patterns are not unique since
PCDM can be used to find them with arbitrary unitary rotation, where
TCM is a particular case of PCDM.
\end{enumerate}
The number of orthogonal basis radiation patterns that a pixel antenna
can effectively excite is referred to as the \textit{effective aerial
degrees of freedom} (EADoF), denoted by $R$ \cite{kalis2008novel}.
It is given by the rank of open-circuit radiation pattern matrix using
SVD, or the rank of the correlation matrix of open-circuit radiation
patterns using PCDM. However, in practice when numerically performing
SVD (or EVD using PCDM), there generally are some very small singular
values or eigenvalues, making it difficult to accurately evaluate
the EADoF. To solve this issue, we can use the cumulative radiated
power to determine EADoF which gives the smallest number of orthogonal
basis radiation patterns that reach the threshold of cumulative power.
This approach retains the dominant basis radiation patterns while
excluding those difficult to excite.

\subsection{Pattern Coder}

\begin{figure}[t]
\begin{centering}
\textsf{\includegraphics[width=8cm]{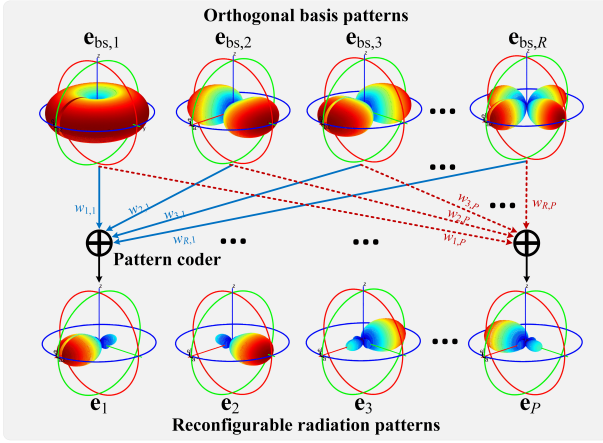}}
\par\end{centering}
\caption{Illustration of pixel antenna radiation pattern synthesis by linearly
coding the orthogonal basis radiation patterns using pattern coders.
\label{fig:Beam}}
\end{figure}
Leveraging the orthogonal basis radiation patterns, we can decompose
the radiation pattern of a pixel antenna $\mathbf{e}$ into a linear
combination of orthogonal basis radiation patterns, written as $\mathbf{e}=\mathbf{E}_{\mathrm{bs}}\mathbf{w}\left(\mathbf{b}\right)$,
where $R$ orthogonal basis radiation patterns are collected into
a matrix $\mathbf{E}_{\mathrm{bs}}=\left[\mathbf{e}_{\mathrm{bs,}1},...,\mathbf{e}_{\mathrm{bs,}R}\right]$
and the coefficients are collected into a vector $\mathbf{w}\left(\mathbf{b}\right)=\left[w_{1},...,w_{R}\right]\in\mathbb{C}^{R\times1}$
referred to as the \textit{pattern coder} \cite{Han2026}. The pattern
coder is a function of antenna coder $\mathbf{b}$ and determines
how these basis patterns are linearly coded to synthesize the radiation
pattern, as shown in Fig. \ref{fig:Beam}.

Now we can further reformulate the beamspace channel model and get
more insights into pixel antennas. For example, in a SISO system where
a pixel antenna is used at the transmitter, we can rewrite the beamspace
channel as $h=\mathbf{h}\mathbf{w}\left(\mathbf{b}\right)$ where
$\mathbf{h}=\mathbf{e}^{T}_{\mathrm{R}}\mathbf{H}_{\mathrm{V}}\mathbf{E}_{\mathrm{bs}}$.
Namely, we can augment the SISO channel between transmit and receive
antennas into a multiple-input single-output (MISO) channel between
transmit and receive orthogonal patterns. Such reformulation is also
beneficial to reducing the channel estimation overhead since we only
need to estimate $\mathbf{h}$ instead of $\mathbf{H}_{\mathrm{V}}$.
In addition, when assuming a rich scattering environment with entries
of $\mathbf{H}_{\mathrm{V}}$ following i.i.d. complex Gaussian distribution,
the augmented channel $\mathbf{h}$ also follows i.i.d. complex Gaussian
distribution, which is equivalent to conventional Rayleigh fading
MISO channel. Moreover, the beamspace channel representation can also
be used to model millimeter wave channel where $\mathbf{H}_{\mathrm{V}}$
is sparse due to the limited paths.

Overall, antenna coding determines current distributions on pixel
antennas and can dynamically shape radiation patterns. This coding
mechanism inherits the fundamental principle of conventional coding
in communication systems while extending its scope from the digital
signal domain into the EM domain. Antenna coding allows the antenna
itself to become a new dimension in wireless system design and optimization,
via adjusting the antenna topology to either adapt dynamic channels
or convey additional information bits. From this perspective, by leveraging
the antenna coding technique, antennas are no longer static transducers
but rather programmable EM interfaces where the EM theory and communication
theory converge into a unified framework.

\section{Classification of Antenna Coding}

\subsection{Binary, Quantized, and Continuous Antenna Coding}

Antenna coding can be classified into three categories, binary, quantized,
and continuous antenna coding, according to the value of $b_{q}$.
Each category offers distinct trade-offs among reconfigurability,
hardware complexity, and switching speed for implementations.

\textbf{Binary Antenna Coding}: In its most fundamental form, pixel
antennas are implemented by using RF switches, such as PIN diodes,
RF MEMS, and CMOS FET. Each switch can be independently set to either
ON (short-circuit) or OFF (open-circuit) state represented by $b_{q}=0$
or $b_{q}=1$ respectively, yielding $2^{Q}$ possible antenna coders.
Binary antenna coding is inherently compatible with digital control
logic where a bit sequence is generated to control RF switch states,
allowing seamless integration with baseband digital processing. Besides,
the switching speed of RF switches is on the order of nanoseconds,
enabling real-time adaptation to fast-fading channels.

\textbf{Quantized Antenna Coding}: The basic binary antenna coding
can be further extended to a more general form where each variable
lumped element supports a finite number of quantized states beyond
simple ON/OFF. Such elements can be implemented by single-pole multi-throw
switches connecting with multiple reactive loads, or by digitally
tunable capacitors which offer a programmable set of discrete capacitive
reactances. In this case, $b_{q}$ can take $L$ quantized values
within $\left[-1,1\right]$ and the total number of possible antenna
coders becomes $L^{Q}$. Therefore, albeit at the cost of higher control
and hardware complexity, quantized antenna coding offers greater reconfigurability
than the binary case.

\textbf{Continuous Antenna Coding}: To achieve finer control over
pixel antennas, continuous antenna coding can be employed where $b_{q}$
can take any real value within $\left[-1,1\right]$, corresponding
to any reactance within $\left[-x_{\mathrm{oc}},x_{\mathrm{oc}}\right]$.
This can be implemented by using varactors to continuously tune the
reactances through varying DC biased voltages. In practice, considering
the finite tuning capability of varactors in capacitive reactance
range, $b_{q}$ can only take values within a subset of $\left[-1,1\right]$,
which reduces the reconfigurablity compared with the ideal continuous
case. Besides, to make the reactance range symmetrically covers both
capacitive and inductive regions, a fixed offset inductor can be introduced
in series with each varactor. However, the continuous antenna coding
significantly enriches the reconfigurability of pixel antennas by
an infinite number of possible antenna coders. This enables smooth
transitions between different radiation patterns when switching antenna
coders. However, varactors require precise biased voltage control
and typically exhibit slower tuning speeds, on the order of microseconds,
which may limit their applicability in fast-varying environments.

Overall, binary antenna coding has the smallest discrete search space
with limited reconfigurability; quantized antenna coding offers more
radiation patterns at higher complexity; continuous antenna coding
has the largest search space with the most flexible reconfigurability
but requires more complex hardware. This trade-off needs to be considered
when selecting proper antenna coding schemes.

\subsection{Slow and Fast Antenna Coding}

Antenna coding can also be classified into slow and fast antenna coding,
depending on the switching interval.

\textbf{Slow Antenna Coding}: It remains constant over a block of
symbols within the channel coherence time and is switched according
to channel state information (CSI) which can be estimated through
beamspace pilot transmission where pilot signals are sent by sequentially
using orthogonal basis radiation patterns. In slow antenna coding,
radiation pattern of a pixel antenna can be optimized to adapt to
the time-varying propagation environment, e.g., constructively combining
multiple paths to enhance the channel gain or deconstructively combining
to suppress the interference. Thus, the system performance such as
channel capacity can be improved compared with conventional antenna
with fixed configuration.

\textbf{Fast Antenna Coding}: The switching interval of antenna coders
is at the symbol level. Fast antenna coding is utilized to modulate
more information bits on the radiation pattern of pixel antennas.
For example, the indices of antenna coders can be mapped into bits.
By selecting a specific antenna coder, the corresponding symbol is
modulated on the radiation pattern and transmitted in the beamspace
domain. Thus, fast antenna coding introduces an extra modulation to
break through SE bound of conventional systems which perform modulation
solely at the baseband. At the receiver, joint estimation on both
the baseband symbol and the antenna coder, with the knowledge of CSI,
can be performed for detection.

\section{Cases Study}

\begin{figure}[t]
\begin{centering}
\textsf{\includegraphics[width=8cm]{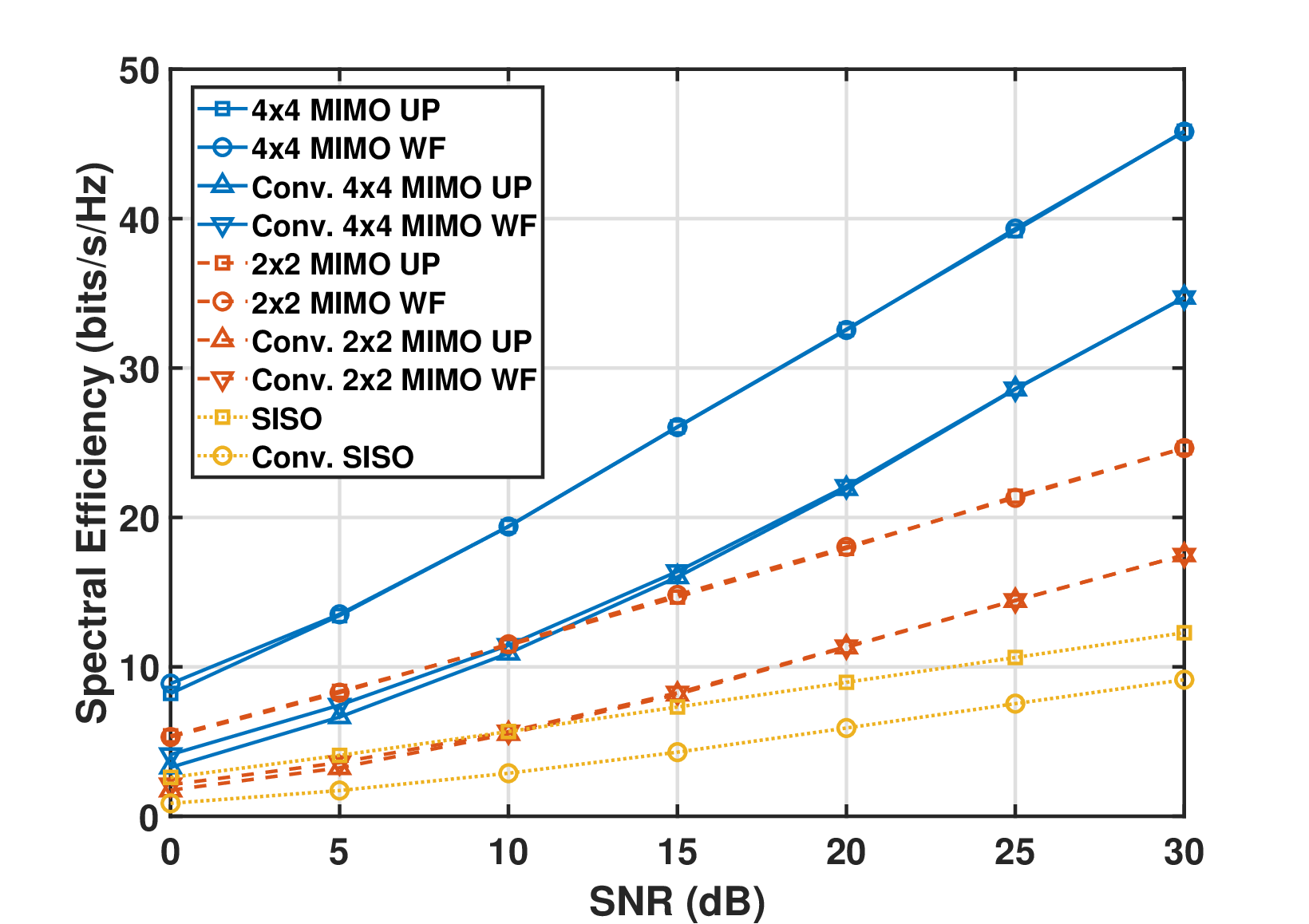}}
\par\end{centering}
\caption{Spectral efficiency of the SISO and MIMO pixel antenna systems with
slow antenna coding. \label{fig:Slow}}
\end{figure}
In this section, we study two cases to demonstrate how antenna coding
unlocks a new dimension for performance enhancement in wireless communications.

\subsection{Channel Gain Enhancement}

Consider a SISO system, where a conventional antenna is used at the
transmitter and a pixel antenna with EADoF of $R$ is used at the
receiver. Using the orthogonal basis radiation patterns, the SISO
channel with pixel antenna is augmented into a 1-by-$R$ single-input
multiple output (SIMO) channel. Thus, by utilizing the maximum ratio
combining (MRC), the average channel gain of the SISO system is upper
bounded by the EADoF $R$. To enhance the channel gain and approach
the upper bound, slow antenna coding can be used, where the antenna
coder is optimized based on CSI \cite{shen2024antenna}.

Further considering a MIMO system with both the transmitter and receiver
employing pixel antennas, the transmit and receive antenna coding
can be jointly optimized to maximize the channel capacity. Under the
uniform power (UP) allocation, only antenna coders need to be optimized.
However, when using the water-filling (WF), the power allocation is
coupled with antenna coders, requiring a joint optimization for antenna
coders and power allocation to maximize capacity.

Fig. \ref{fig:Slow} shows the performance of the SISO and MIMO systems
with slow antenna coding. It is shown that SE is significantly enhanced
by using slow antenna coding compared with conventional systems with
fixed antenna configurations, since radiation patterns of pixel antennas
can be flexibly adjusted to adapt to the channel and enhance the channel
gain.

\subsection{Multiplexing Gain Enhancement}

\begin{figure}[t]
\begin{centering}
\textsf{\includegraphics[width=8cm]{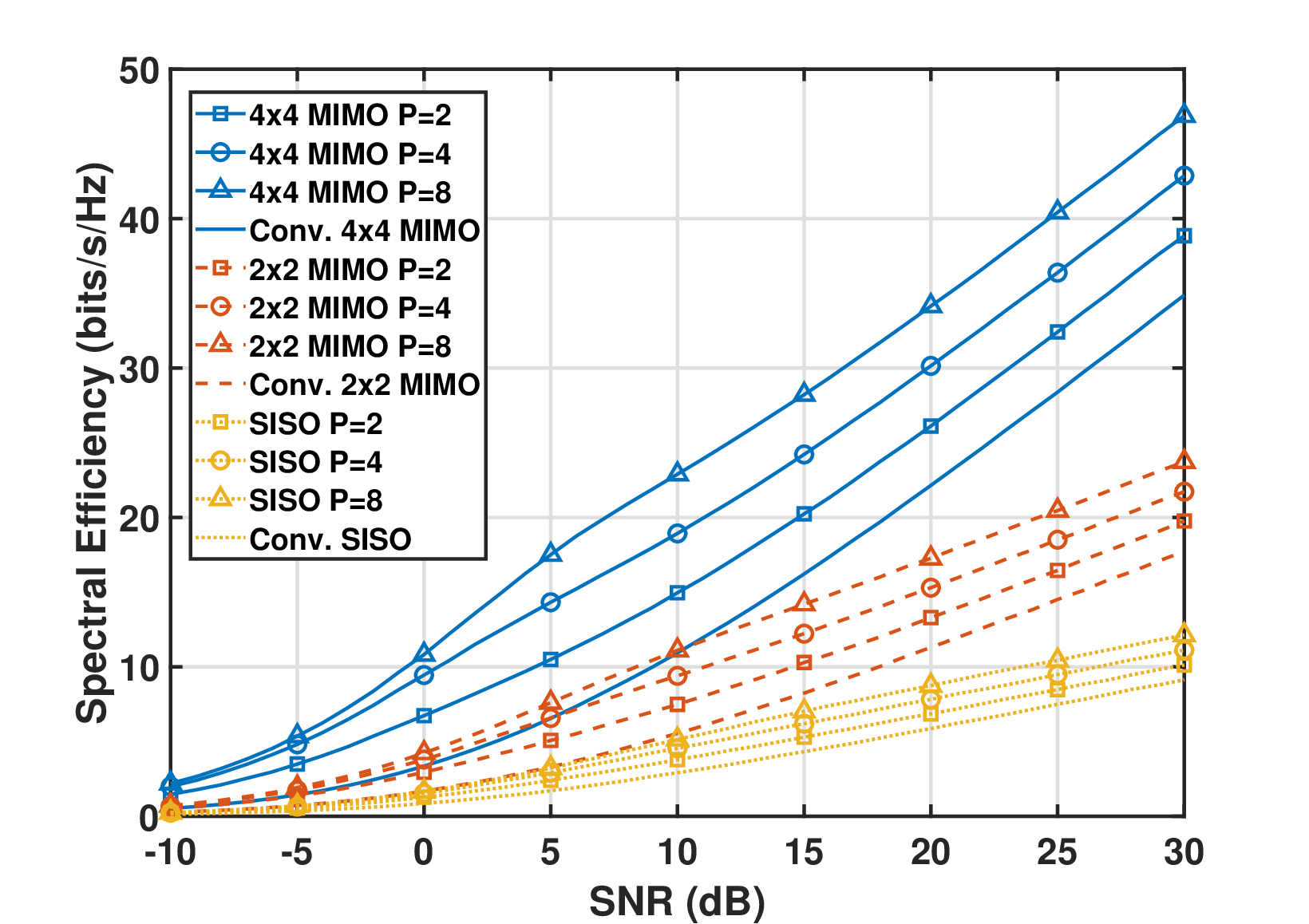}}
\par\end{centering}
\caption{Spectral efficiency of the SISO and MIMO pixel antenna systems with
fast antenna coding. \label{fig:Fast}}
\end{figure}
Considering a MIMO system employing pixel antennas at the transmitter
and conventional antennas at the receiver, fast antenna coding can
be performed by synchronizing antenna coder switching with the baseband
symbol period \cite{Han2026}. Each pixel antenna selects one of the
$P$ antenna coders to transmit additional information bits. Therefore,
for each symbol, $\mathrm{log}_{2}P$ bits are carried by the radiation
pattern of the pixel antenna while $\mathrm{log}_{2}M$ bits are carried
by the conventional symbol with $M$-ary modulation. The total number
of bits per symbol transmitted by a pixel antenna with fast antenna
coding is $\mathrm{log}_{2}P+\mathrm{log}_{2}M$, which is $\mathrm{log}_{2}P$
bits more than conventional antenna with fixed configuration. This
is achieved by exploiting spatial multiplexing gain using fast antenna
coding, instead of increasing bandwidth or RF chains. However, a trade-off
between multiplexing gain and bit error rate (BER) needs to be considered,
where a larger $P$ is beneficial to enhance SE at high signal-to-noise
ratio (SNR) while $P$ should be reduced to maintain acceptable BER
at low SNR.

A direct design of fast antenna coding is to modulate information
bits directly onto orthogonal basis radiation patterns. That is, $P$
antenna coders are used to excite $P$ orthogonal pattern coders $\left(P\leqslant R\right)$.
Fig. \ref{fig:Fast} shows the performance of the SISO and MIMO systems
with fast antenna coding. It can be observed that compared with conventional
systems, SE can be significantly enhanced by using fast antenna coding.
Particularly, SE of the $N$-by-$N$ MIMO system is improved by $N\mathrm{log}_{2}P$
bits/s/Hz at high SNR, breaking through SE of conventional MIMO systems.
The multiplexing gain can be further improved by using more radiation
patterns to transmit information bits, even if these radiation patterns
are not mutually orthogonal.

\section{Optimization Techniques}

While pixel antennas offer high reconfigurablity, it remains challenging
to efficiently optimize antenna coding over a high-dimensional space,
particularly for binary antenna coding optimization. Thus, optimization
techniques are critical to determine the performance of pixel antenna
systems.

\subsection{Successive Exhaustive Boolean Optimization}

Successive exhaustive Boolean optimization (SEBO) reduces the complexity
of binary optimization by dividing the antenna coder into smaller
blocks and sequentially optimizing each block while fixing the other
blocks \cite{7762757}. Specifically, the antenna coder is firstly
partitioned into small blocks. The algorithm optimizes each block
at a time by exhaustively searching all possible combinations within
that block while the remaining blocks are fixed. This block-wise search
repeats cyclically until convergence. Then a random perturbation step
is introduced to escape local optima. The algorithm randomly flips
a few bits and checks whether the objective improves. If no improvement
is found, the number of flipped bits is gradually increased, until
it increases to the block size, and the algorithm continues. A key
advantage of SEBO is the simplicity since only one parameter, the
maximum block size, is set in SEBO.

\subsection{Evolutionary Algorithms}

Evolutionary algorithms maintain a population of antenna coder candidates
that evolve over generations to find high-quality solutions. Representative
algorithms include genetic algorithm (GA) and binary particle swarm
optimization (BPSO). Evolutionary algorithms can approach the global
optimum at the cost of higher computational complexity. Therefore,
it is suitable to use evolutionary algorithms to find high-quality
antenna coders for offline training such as in the codebook design
or supervised learning as detailed below. In addition, evolutionary
algorithms require careful tuning parameters, such as crossover rate,
mutation rate of GA, and inertia weight, acceleration coefficients
of BPSO.

\subsection{Codebook-Based Methods}

Codebook-based methods are based on evaluating each codeword in a
pre-designed codebook and selecting the best codeword, which reduces
the computational complexity to linear level. Thus, codebook design
critically determines performance of codebook-based methods. Typically,
the codebook is designed offline based on a training set of channel
realizations by using the generalized Lloyd algorithm (GLA). In GLA,
partition optimization, which optimizes the partition of training
set associated with each codeword, and centroid optimization, which
optimizes the codeword for each partition using either SEBO or evolutionary
algorithms, are alternatively performed until convergence. A key issue
in the codebook design is choosing the codebook size, as a larger
codebook offers better performance at the cost of higher search complexity.
To alleviate this issue, hierarchical codebook can be designed to
accelerate the searching process. Once designed, the codebook can
be used across various deployments in similar environments, making
codebook-based methods practical and efficient.

\subsection{Machine Learning and AI-Based Methods}

Machine learning and AI-based methods can train models to instantly
generate antenna coders, which is much faster than aforementioned
methods. The computational complexity is shifted from online searching
to offline training. Among different learning algorithms, supervised
deep learning is straightforward for antenna coding design. The training
data pairs each input (CSI or transmit symbol) with its corresponding
antenna coder (optimized by SEBO or evolutionary algorithms) as the
label. A deep neural network is then trained to learn the mapping
from input to label. Once trained, the network can be widely deployed
in real time.

In summary, SEBO and codebook-based methods are recommended when low-latency
online optimization is required. Evolutionary algorithms are more
suitable when performance is the primary concern and computational
time is not constrained. AI-based methods are preferred for real-time
deployment in time-varying environment.

\section{Challenges and Opportunities}

Despite the promising capabilities of antenna coding, several practical
challenges remain to be solved. In this section, we discuss these
challenges and opportunities for future works.

\subsection{Hardware Complexity}

The hardware complexity of pixel antennas increases as the number
of variable lumped elements $Q$ becomes larger, since each variable
lumped element requires a DC biasing circuit. To address this challenge,
we can optimize antenna coding with the goal of reducing $Q$ while
maintaining good system performance. For example, by identifying pixel
ports where all antenna coders share the same value, the corresponding
variable lumped elements can be replaced by fixed loads. If no such
port exists, forcing one pixel port to have the same value across
all antenna coders with least influence and optimizing the remaining
pixel ports can be performed to meet the goal.

\subsection{Non-Ideal Variable Lumped Elements}

The variable lumped elements introduce insertion loss due to parasitic
resistances, leading to degraded radiation efficiency and system performance.
Using variable lumped elements with low parasitic resistances and
reducing the number of variable lumped elements can mitigate this
issue. In addition, minimizing the insertion loss can be incorporated
into antenna coding optimization by modeling parasitic resistances
in the multiport network model.

\subsection{Impedance Matching}

The input impedance of a pixel antenna varies when antenna coder is
changed, resulting in impedance mismatching between the RF chain and
pixel antenna, thus degrading the system performance. One solution
is to use a dynamic matching network tuned by RF switches or varactors.
This matches the input impedance of a pixel antenna with different
antenna coders to the RF chain in real time. Another approach is to
incorporate the input impedance into antenna coding optimization,
for example, by including the reflection coefficient as part of the
objective function, so that antenna coders that cause severe mismatching
can be excluded.

\subsection{Mutual Coupling}

In the MIMO pixel antenna system, each pixel antenna is connected
to an RF chain. The mutual coupling among pixel antennas distorts
radiation patterns and degrades performance. Antenna decoupling techniques
such as inserting passive decoupling structures between pixel antennas
are effective to suppress mutual coupling. Besides, mutual coupling
can be considered in antenna coding optimization by extending multiport
network model to multiple pixel antennas, so that antenna coders for
all pixel antennas will be jointly optimized with the consideration
of mutual coupling.

\subsection{Power Consumption}

PIN diodes require forward bias currents to maintain ON states so
that the power consumption (e.g., typically 20 mW per diode) scales
with the number of active PIN diodes. Alternative solution is using
MEMS switches which are actuated by voltage with low power consumption,
e.g., typically 10 $\mathrm{\mu}$W. Besides, varactors draw negligible
DC current under reverse bias, resulting in negligible power consumption.

\section{Conclusion}

Antenna coding technique leverages the new DoF provided by antenna
reconfigurability to break through the performance bound of conventional
wireless communication systems. Utilizing high reconfigurability of
pixel antennas, radiation patterns can be adjusted by switching antenna
coders so as to convey additional information bits or match the propagation
channel in beamspace. The choice of antenna coding scheme fundamentally
determines the achievable system performance, which depends on the
specific system objectives and hardware constraints.

\bibliographystyle{IEEEtran}

\begin{thebibliography}{10}
	\providecommand{\url}[1]{#1}
	\csname url@samestyle\endcsname
	\providecommand{\newblock}{\relax}
	\providecommand{\bibinfo}[2]{#2}
	\providecommand{\BIBentrySTDinterwordspacing}{\spaceskip=0pt\relax}
	\providecommand{\BIBentryALTinterwordstretchfactor}{4}
	\providecommand{\BIBentryALTinterwordspacing}{\spaceskip=\fontdimen2\font plus
		\BIBentryALTinterwordstretchfactor\fontdimen3\font minus
		\fontdimen4\font\relax}
	\providecommand{\BIBforeignlanguage}[2]{{%
			\expandafter\ifx\csname l@#1\endcsname\relax
			\typeout{** WARNING: IEEEtran.bst: No hyphenation pattern has been}%
			\typeout{** loaded for the language `#1'. Using the pattern for}%
			\typeout{** the default language instead.}%
			\else
			\language=\csname l@#1\endcsname
			\fi
			#2}}
	\providecommand{\BIBdecl}{\relax}
	\BIBdecl
	
	\bibitem{1367557}
	B.~Cetiner, H.~Jafarkhani, J.-Y. Qian, H.~J. Yoo, A.~Grau, and F.~De~Flaviis,
	``Multifunctional reconfigurable {MEMS} integrated antennas for adaptive
	{MIMO} systems,'' \emph{IEEE Commun. Mag.}, vol.~42, no.~12, pp. 62--70,
	2004.
	
	\bibitem{11302793}
	T.~Wu, K.~Zhi, J.~Yao, X.~Lai, J.~Zheng, H.~Niu, M.~Elkashlan, K.-K. Wong,
	C.-B. Chae, Z.~Ding, G.~K. Karagiannidis, M.~Debbah, and C.~Yuen, ``Fluid
	antenna systems enabling 6{G}: Principles, applications, and research
	directions,'' \emph{IEEE Wirel. Commun.}, vol.~33, no.~4, pp. 100--108, 2026.
	
	\bibitem{11513728}
	Y.~Tian, J.~Ren, T.~Wu, W.~Liu, M.~Elkashlan, M.~C. Valenti, N.~Al-Dhahir, and
	H.~C. So, ``Hybrid architecture gets fluid: A new paradigm for
	direction-of-arrival estimation in 6{G} networks,'' \emph{IEEE Trans. Signal
		Process.}, vol.~74, pp. 2083--2099, 2026.
	
	\bibitem{11263876}
	K.-K. Wong, C.~Wang, S.~Shen, C.-B. Chae, and R.~Murch, ``Reconfigurable pixel
	antennas meet fluid antenna systems: A paradigm shift to electromagnetic
	signal and information processing,'' \emph{IEEE Wirel. Commun.}, vol.~33,
	no.~1, pp. 191--198, 2026.
	
	\bibitem{shen2024antenna}
	S.~Shen, K.-K. Wong, and R.~Murch, ``Antenna coding empowered by pixel
	antennas,'' \emph{IEEE Trans. Commun.}, vol.~74, pp. 446--460, 2026.
	
	\bibitem{Han2026}
	Z.~Han, S.~Shen, and R.~Murch, ``Exploiting spatial multiplexing based on pixel
	antennas: An antenna coding approach,'' \emph{IEEE J. Sel. Topics Signal
		Process.}, vol.~20, no.~3, pp. 233--246, 2026.
	
	\bibitem{lotfi2017printed}
	P.~Lotfi, S.~Soltani, and R.~D. Murch, ``Printed endfire beam-steerable pixel
	antenna,'' \emph{IEEE Trans. Antennas Propag.}, vol.~65, no.~8, pp.
	3913--3923, 2017.
	
	\bibitem{Zhang2022a}
	Y.~Zhang, S.~Tang, Z.~Han, J.~Rao, S.~Shen, M.~Li, C.-Y. Chiu, and R.~Murch,
	``A low-profile microstrip vertically polarized endfire antenna with 360
	beam-scanning and high beam-shaping capability,'' \emph{IEEE Trans. Antennas
		Propag.}, vol.~70, no.~9, pp. 7691--7702, 2022.
	
	\bibitem{Zhang2022}
	Y.~Zhang, Z.~Han, S.~Tang, S.~Shen, C.-Y. Chiu, and R.~Murch, ``A highly
	pattern-reconfigurable planar antenna with 360 single- and multi-beam
	steering,'' \emph{IEEE Trans. Antennas Propag.}, vol.~70, no.~8, pp.
	6490--6504, 2022.
	
	\bibitem{6678331}
	S.~Song and R.~D. Murch, ``An efficient approach for optimizing frequency
	reconfigurable pixel antennas using genetic algorithms,'' \emph{IEEE Trans.
		Antennas Propag.}, vol.~62, no.~2, pp. 609--620, 2014.
	
	\bibitem{sayeed2002deconstructing}
	A.~M. Sayeed, ``Deconstructing multiantenna fading channels,'' \emph{IEEE
		Trans. Signal Process.}, vol.~50, no.~10, pp. 2563--2579, 2002.
	
	\bibitem{mautz1973modal}
	J.~Mautz and R.~Harrington, ``Modal analysis of loaded {N}-port scatterers,''
	\emph{IEEE Trans. Antennas Propag.}, vol.~21, no.~2, pp. 188--199, 1973.
	
	\bibitem{PCDM}
	Z.~Han, S.~Shen, Y.~Zhang, C.-Y. Chiu, and R.~Murch, ``A pattern correlation
	decomposition method for analysis of {ESPAR} in single-{RF} {MIMO} systems,''
	\emph{IEEE Trans. Wirel. Commun.}, vol.~21, no.~7, pp. 4654--4668, 2022.
	
	\bibitem{kalis2008novel}
	A.~Kalis, A.~G. Kanatas, and C.~B. Papadias, ``A novel approach to {MIMO}
	transmission using a single {RF} front end,'' \emph{IEEE J. Sel. Areas
		Commun.}, vol.~26, no.~6, pp. 972--980, 2008.
	
	\bibitem{7762757}
	S.~Shen, Y.~Sun, S.~Song, D.~P. Palomar, and R.~D. Murch, ``Successive boolean
	optimization of planar pixel antennas,'' \emph{IEEE Trans. Antennas Propag.},
	vol.~65, no.~2, pp. 920--925, 2017.
	
\end{thebibliography}

$\vphantom{}$

{\footnotesize\textbf{Zixiang Han}}{\footnotesize{} {[}M{]} is currently
a Senior Research Scientist and Project Manager with Future Research
Laboratory, China Mobile Research Institute, Beijing, China. He received
the 6G Young Scholar Award at the 2024 Global 6G Conference.}{\footnotesize\par}

{\footnotesize\textbf{Shanpu Shen}}{\footnotesize{} {[}SM{]} is currently
an Associate Professor with the State Key Laboratory of Internet of
Things for Smart City and Department of Electrical and Computer Engineering,
University of Macau, Macau. He is serving as an Associate Editor for
IEEE Transactions on Antennas and Propagation.}{\footnotesize\par}

{\footnotesize\textbf{Yifei Yuan}}{\footnotesize{} {[}F{]} is currently
a Chief Expert with China Mobile Research Institute, Beijing, China,
responsible for 6G study. He has extensive publications, including
14 books on LTE-Advanced, 5G and 6G. He has over 60 granted US patents.}{\footnotesize\par}

{\footnotesize\textbf{Jiangzhou Wang}}{\footnotesize{} {[}F{]} is a
Professor at Southeast University, China. He is an International Member
of the Chinese Academy of Engineering (CAE), a Fellow of the Royal
Academy of Engineering (RAEng), U.K., and Fellow of the IEEE. He is
the General Chair of the IEEE GLOBECOM2026, Macau.}{\footnotesize\par}

{\footnotesize\textbf{Ross Murch}}{\footnotesize{} {[}F{]} is a Chair
Professor with the Department of Electronic and Computer Engineering,
Hong Kong University of Science and Technology. His research focus
is creating new RF wave technology for making a better world. He is
a fellow of IEEE, IET and HKIE.}{\footnotesize\par}

{\footnotesize\textbf{Chih-Lin I}}{\footnotesize{} {[}LF{]} is currently
the Chief Scientist, China Mobile, Beijing, China. She is spearheading
major initiatives including 6G, O-RAN, high energy efficiency system
architectures, technologies, and devices, and green energy.}{\footnotesize\par}
\end{document}